\documentclass[%
onecolumn,pra,
superscriptaddress,
nofootinbib,
 amsmath,amssymb,
 aps,
11pt,
]{revtex4-1}

\usepackage[english]{babel}
\usepackage[utf8x]{inputenc}
\usepackage[T1]{fontenc}

\usepackage[a4paper,top=3cm,bottom=2cm,left=3cm,right=3cm,marginparwidth=1.75cm]{geometry}

\usepackage{graphicx}
\usepackage[colorinlistoftodos]{todonotes}
\usepackage[colorlinks=true, allcolors=blue]{hyperref}

\usepackage{braket}

\usepackage{xcolor}
\def\note#1{}

\newtheorem{theorem}{Theorem}

\begin{document}

\title{Quantum Neuron: an elementary building block for machine learning on quantum computers}

\author{Yudong Cao}
\email{yudongcao@fas.harvard.edu}
\affiliation{Department of Chemistry and Chemical Biology, Harvard University, Cambridge, MA 02138}
\author{Gian Giacomo Guerreschi}
\email{gian.giacomo.guerreschi@intel.com}
\affiliation{Parallel Computing Lab, Intel Corporation, Santa Clara, CA 95054}
\author{Al\'{a}n Aspuru-Guzik}
\email{alan@aspuru.com}
\affiliation{Department of Chemistry and Chemical Biology, Harvard University, Cambridge, MA 02138}
\affiliation{Senior Fellow, Canadian Institute for Advanced Research, Toronto, Ontario M5G 1Z8, Canada}


\begin{abstract}

Even the most sophisticated artificial neural networks
are built by aggregating substantially identical units called neurons. A neuron receives multiple signals, internally combines them, and applies a non-linear function to the resulting weighted sum. Several attempts to generalize neurons to the quantum regime have been proposed, but all proposals collided with the difficulty of implementing non-linear activation functions, which is essential for classical neurons, due to the linear nature of quantum mechanics.
Here we propose a solution to this roadblock in the form of a small quantum circuit that naturally simulates neurons with threshold activation. Our quantum circuit defines a building block, the ``quantum neuron'', that can reproduce a variety of classical neural network constructions while maintaining the ability to process superpositions of inputs and preserve quantum coherence and entanglement.
In the construction of feedforward networks of quantum neurons, we provide numerical evidence that the network not only can learn a function when trained with superposition of inputs and the corresponding output, but that this training suffices to learn the function on all individual inputs separately.
When arranged to mimic Hopfield networks, quantum neural networks exhibit properties of associative memory. Patterns are encoded using the simple Hebbian rule for the weights and we demonstrate attractor dynamics from corrupted inputs.
Finally, the fact that our quantum model closely captures (traditional) neural network dynamics implies that the vast body of literature and results on neural networks becomes directly relevant in the context of quantum machine learning.
\end{abstract}

\maketitle

Machine learning systems are revolutionizing the field of data analysis and pattern recognition. Their commercial deployment has already generated a very concrete change in many activities of the everyday life such as travel booking \cite{Booking16}, navigation \cite{MITSelfDriving}, media recommendation \cite{DNNYoutube}, image recognition \cite{Krizhevsky:2017:ICD:3098997.3065386} and playing competitive board games \cite{Silver2016}. Much of the rapid development is driven by deep neural networks, together with recent improvements of training techniques and the availability of massive amounts of data and computational power. Despite their diversity, all neural networks are essentially based on a common unit from which they derive their name: the artificial neuron.

Inspired by the basic mechanism involved in neural activities of living organisms \cite{McCulloch1943Activity,Rabinovich2006DynamicalNeuroscience}, the artificial neurons are organized in networks where the output of one neuron constitutes the inputs for other neurons. Typically, every neuron combines the input values through a weighted sum, applies a non-linear activation function and produces the corresponding value as output. The activation function often takes the form of a step function or, in modern uses, of a continuous sigmoid function. Its non-linearity is an essential feature that makes the collective dynamics dissipative and attractor-based \cite{Rabinovich2006DynamicalNeuroscience,Hopfield1982NeuralAbilities} and contributes to the ability of neural networks to capture highly non-trivial patterns \cite{Hinton2006ReducingNetworks,Hopfield1982NeuralAbilities}.

Independently, the advent of quantum computation \cite{Nielsen2000QuantumInformation} has provided an entirely new perspective for thinking about how information is stored and manipulated. By taking advantage of uniquely quantum mechanical features such as superposition and entanglement, hard computational tasks can be solved with quantum computers significantly faster compared to the best known classical algorithms \cite{Shor1997Polynomial-TimeComputer,Grover1996ASearch,Harrow2009QuantumEquations}. Together with the development of universal quantum computation, an active line of inquiry has appeared in the literature regarding ``quantum neural networks'' (QNN) \cite{Schuld2014TheNetwork}, namely devices or algorithms which combine the unique features of both quantum mechanics and neural networks to perform meaningful computational tasks.

Current proposals of QNN involve a rather diverse collection of ideas with varying degrees of proximity to classical neural networks \cite{Narayanan:2000:QAN:361352.361359,Schuld2014TheNetwork,Kak1995OnComputing,Zak1998QuantumNets,Behrman2000SimulationsNetworks,Kak1995OnComputing,Wan2016QuantumNetworks,Ventura1998QuantumCapacity,Rebentrost2017ANetwork}. The central issue in QNN lies in the problem of incorporating the non-linear, dissipative dynamics of classical neural networks into the linear, unitary framework of quantum mechanics. Potential resolutions attempted so far include introducing quantum measurements \cite{Kak1995OnComputing,Zak1998QuantumNets}, exploiting the quadratic form of kinetic term to generate non-linearity \cite{Behrman2000SimulationsNetworks}, using dissipative quantum gates \cite{Kak1995OnComputing} and reversible circuits \cite{Wan2016QuantumNetworks}. Other proposals are successful in capturing certain aspects of classical neural networks such as the associative memory property \cite{Ventura1998QuantumCapacity,Rebentrost2017ANetwork}, but deviate in fundamental ways from classical neural networks. Recent reviews \cite{Schuld2014TheNetwork,Ciliberto2017a} acknowledge the lack of a construction that fully incorporates both the unique properties of quantum mechanics and the nonlinear features of neural networks.

\begin{figure}
\includegraphics[scale=0.97]{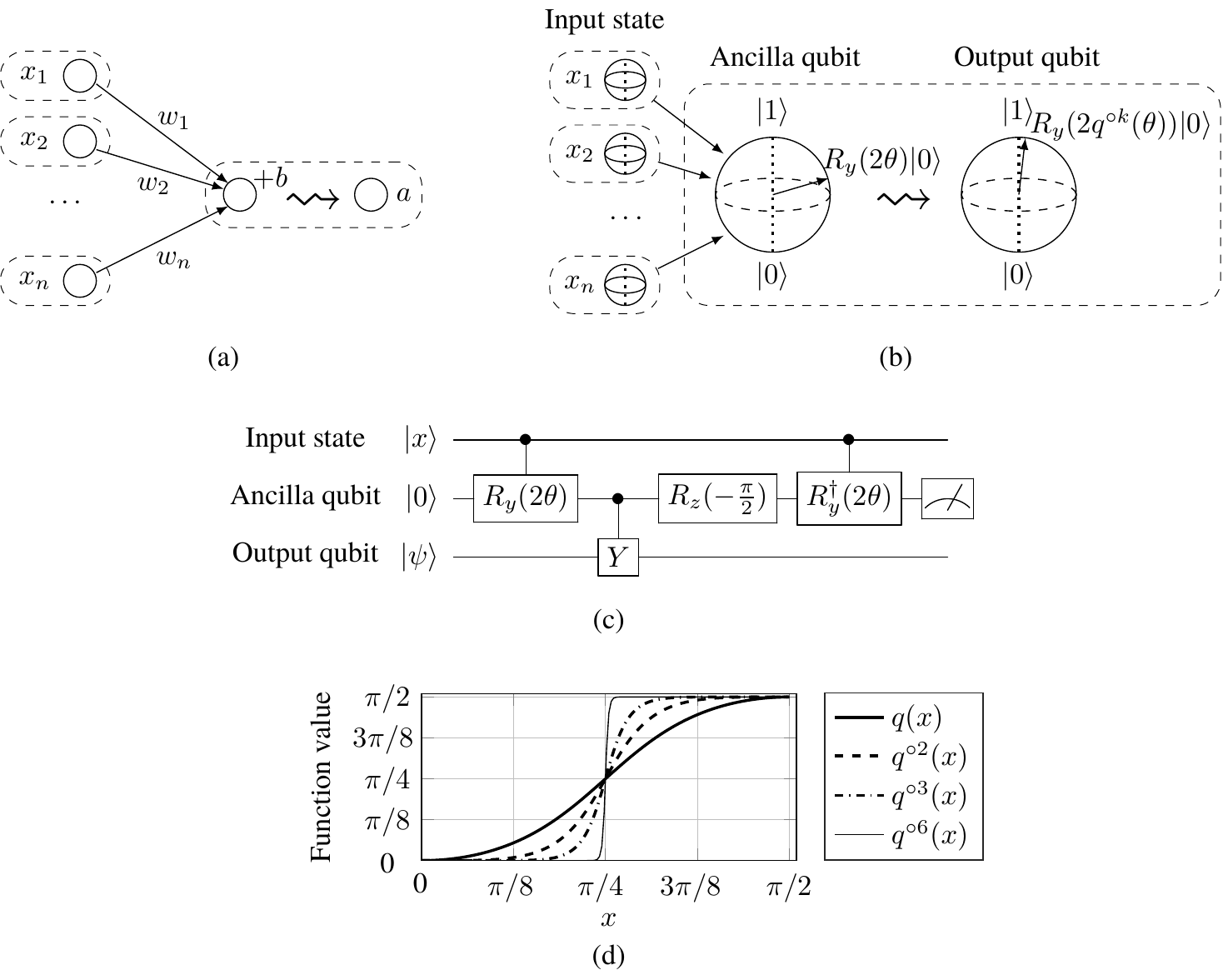}
\caption{Basic setup of the quantum neuron model.
\textbf{(a) The classical neuron} (marked using dashed boxes). The inputs $x_1$, $\cdots$, $x_n$ are combined with specific weights $w_i$, and biased by $b$ to form $\theta=w_1x_1+\cdots+w_nx_n+b$. The output activation is $a=\sigma(\theta)$, with $\sigma$ being a sigmoid or step function. 
\textbf{(b) The quantum neuron} (marked using dashed boxes). Bloch sphere visualization of the output qubit state before and after the RUS, corresponding to the linear and non-linear activation function respectively. The function $q$ is shown in subplot d. The notation $\leadsto$ represents the internal update of the neuron corresponding to the activation function. 
In practice it is the ability to use rotations by $2\theta$ to implement rotations by $2q^{\circ k}(\theta)$ via repeat-until-success circuits. The input state is assumed to be prepared by some external method, possibly controlled by other quantum neurons. 
\textbf{(c) Repeat-until-success (RUS) circuit} for realizing rotation with an angle $q(\varphi)=\arctan(\tan^2\varphi)$. Here we use the convention $R_p(\varphi)=\exp(-i P \varphi/2)$ where $p\in\{x,y,z\}$ labels Pauli operators $P\in\{X,Y,Z\}$.
\textbf{(d) Nonlinear function} $q(\varphi)=\arctan(\tan^2 \varphi)$ and its self composition $q^{\circ k}(\varphi)=\arctan(\tan^{2^k} \varphi)$. 
}
\label{fig:basic_setting}
\end{figure}

Here we present the realization of a \emph{quantum neuron} and demonstrate its application as a building block of quantum neural networks. Our approach uses the recently developed repeat-until-success techniques for quantum gate synthesis \cite{Wiebe2013FloatingSynthesis,Paetznick2013Repeat-Until-Success:Unitaries,Wiebe2014QuantumCircuits,Bocharov2015EfficientCircuits}. We show that our model is able to simulate classical neurons with sigmoid or step function activation while processing inputs in quantum superposition.
We describe the design and simulate the performance of classifiers and associative memories in the quantum regime.
In fact, in the context of feedforward neural networks, our model can simulate a standard feedforward network and process all the training data at once in quantum superposition. 
For Hopfield networks \cite{Hopfield1982NeuralAbilities}, we show numerically that our model reproduces the attractor dynamics by converging to a unique, memorized, pattern even when starting from a quantum superposition of input states. 
Our quantum neuron model is the first explicit construction that satisfies the criteria for a reasonable quantum neural network proposed by \cite{Schuld2014TheNetwork} in a way that naturally combines the unique features of both quantum mechanics and machine learning.

$\quad$\\
\noindent{\bf Construction of the quantum neuron}

For the purposes of this work, a classical neuron is a function that takes $n$ variables $x_1$, $x_2$, $\cdots$, $x_n$ and maps them to the output value $a=\sigma(w_1x_1+w_2x_2+\cdots+w_nx_n+b)$ with $\{w_i\}$ and $b$ being the synaptic weights and bias, respectively (Figure~\ref{fig:basic_setting}a). The quantity $\theta=w_1x_1+\cdots+w_nx_n+b$ is called the input signal to the neuron. The activation function $\sigma(z)$ is a nonlinear function. An example of some activation functions considered in classical implementations are the step function, that returns $1$ if $z>0$ and $-1$ otherwise, or continuous functions with ``softer'' nonlinearity, such as the sigmoid function $\text{tanh}(z)=\frac{e^z-e^{-z}}{e^z+e^{-z}}$, or other kinds of nonlinear functions. In all cases, we say that the output value $a\in[-1,1]$ is the \emph{state} of the neuron.

To map this setting to the quantum framework, we introduce a qubit whose quantum state is
$R_y(a\frac{\pi}{2}+\frac{\pi}{2})\ket{0} = \cos(a\frac{\pi}{4}+\frac{\pi}{4})\ket{0} + \sin(a\frac{\pi}{4}+\frac{\pi}{4}) \ket{1}$, where $a\in[-1,1]$ is a scalar and $R_y(t)=\exp(-itY/2)$ is a quantum operation corresponding to the rotation generated by the Pauli Y operator%
\footnote{The presence of the factor $1/2$ inside the expression for $R_y(t)$ has two reasons: First, it is customary in the quantum computing community to include such factor and call this specific operation ``a rotation by angle $t$ around the Y axis of the qubit''. Second it allows for a direct visualization of this operation on quantum states in terms of the Bloch sphere hinted in Figure~\ref{fig:basic_setting}b: $R_y(t)$ effectively rotates the arrow representing the qubit state by an angle $t$. The notational price to pay for such benefits is that most expressions in the following appears like $R_y(2w_i)$ or $R_y(2b)$, with an explicit factor 2 to balance the intrinsic factor $1/2$.}
(see Figure~\ref{fig:basic_setting}b).
The extremal cases $a=-1$ and $a=1$ correspond to quantum states $\ket{0}$ and $\ket{1}$ respectively, in analogy to the classical cases with binary output. 
However, the case $a\in\left(-1,1\right)$ represents the quantum neuron in a superposition of $\ket{0}$ and $\ket{1}$, which has no classical analogy.

\begin{figure}
\hspace{0.15in}
\includegraphics[scale=0.95]{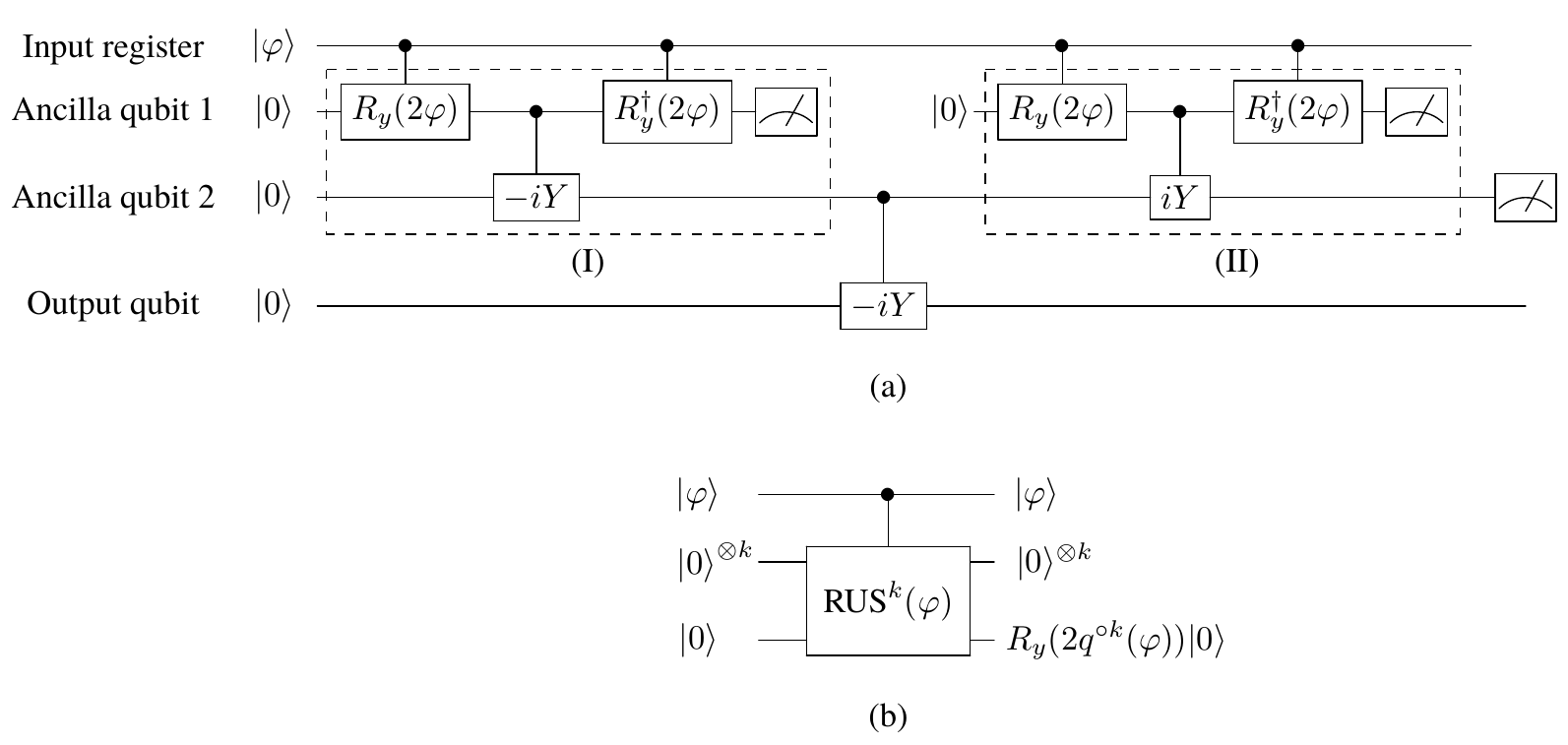}
\caption{
\textbf{(a)} 
{\bf Two iterations of the RUS circuit} shown in Figure \ref{fig:basic_setting}c. Here we have absorbed the $R_z$ rotation in Figure \ref{fig:basic_setting}c into an additional phase of the controlled $Y$ operation (equivalent apart from an unobservable global phase).
\textbf{(b)} {\bf General $k$-iteration RUS circuit}. Here the circuit in (a) can be considered as a special case $k=2$.}
\label{fig:rus_twice}
\end{figure}

\begin{figure}
\begin{center}
\includegraphics[scale=1]{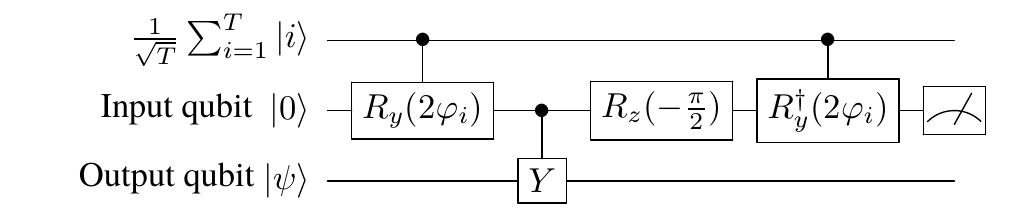}
\end{center}
\caption{RUS applied with a superposition of input rotations. 
}
\label{fig:rus_super}
\end{figure}

In order to mimic, using a quantum circuit, the function of the classical neuron where inputs $x_1,\cdots,x_n\in\{0,1\}$ are linearly combined to form an input $\theta=w_1x_1+\cdots+w_nx_n+b$, one could simply use the state $|x\rangle=|x_1\cdots x_n\rangle$ as a control state and apply $R_y(2w_i)$ onto an ancilla qubit conditioned on the $i$-th qubit, followed by $R_y(2b)$ on the ancilla qubit. This amounts to applying the $R_y(2\theta)$ on the ancilla qubit conditioned on the state $|x\rangle$ of the input neurons (Figure \ref{fig:basic_setting}b and \ref{fig:basic_setting}c). The second step is to perform a rotation by $R_y(2\sigma(\theta))$ 
where $\sigma$ is a non-linear function (either sigmoid or threshold function). We approximate such rotation by a class of circuits called \emph{repeat-until-success (RUS) circuits} \cite{Wiebe2013FloatingSynthesis}. Figure \ref{fig:basic_setting}c shows a circuit which implements $R_y(2q(\theta))$ 
where $q(\theta)=\arctan(\tan^2\theta)$ is a sigmoid-like non-linear function (Figure \ref{fig:basic_setting}d). The action of an RUS circuit on the output qubit depends on the measurement outcome of the ancilla qubit. If the measurement returns $|0\rangle$, this indicates that the rotation by $2q^{\circ k}(\theta)$ has been successfully applied to the output qubit. Otherwise if the ancilla qubit measures $|1\rangle$, this indicates that the circuit has implemented a rotation $R_y(\pi/2)$ onto the output qubit. In this case we correct the operation by applying $R_y(-\pi/2)$ and then repeat the circuit until $|0\rangle$ is measured in the ancilla qubit, hence the name \emph{repeat until success}. 

Using the basic RUS circuit (Figure \ref{fig:basic_setting}c) as a building block, we could realize rotation $R_y(2q^{\circ k}(\theta))$ by recursively applying the basic construction (Figure \ref{fig:rus_twice}). The goal of using RUS is to realize a form of threshold behaviour on $\theta$: if $\theta>\pi/4$ then we would like the output qubit to be as close to $R_y(\pi)|0\rangle=|1\rangle$ as possible and if $\theta<\pi/4$ we would like the output qubit to be as close to $R_y(0)|0\rangle$ as possible. Such threshold behaviour is a key ingredient in realizing neural computation using quantum mechanics \cite{Schuld2014TheNetwork}. We could think of each iteration of RUS as moving the input angle $\theta$ closer and closer to its attractor, which is $0$ or $\pi/2$ depending if $\theta$ is greater than the threshold $\pi/4$. The closer $\theta$ is to the threshold, naturally more iterations are needed for bringing it close to its attractor. In order for an input angle $\theta$ which is at distance at most $\delta$ away from the threshold $\pi/4$, to get it to be at most $\epsilon$ away from the attractor one needs $k=O(\log\frac{1}{\delta\epsilon})$ RUS iterations (see Appendix \ref{subsec:mapq}). The runtime of RUS circuits depends on the history of successes and failures at each measurement. On average the circuit depth scales as $O(14^k)$ for $k$ iterations (Appendix \ref{sec:rus}).

Another feature of RUS circuits is that it can be applied in a quantum superposition. Consider the circuit in Figure \ref{fig:rus_super}, where the input is controlled by a $T$-dimensional register. The controlled rotation onto the input qubit can be written as $\sum_{i=1}^T\ket{i}\bra{i} \otimes R_y(2\varphi_i).$ 
With the control register initialized to a uniform superposition, conditioned on measurement outcome being 0 we have the final state $\frac{1}{\sqrt{T}}\sum_{i=1}^TF_i\ket{i}\otimes\ket{0}\otimes R_y(2q(\varphi_i))\ket{\psi}$ 
which is also a superposition of rotated output states.
{Observe the factor $F_i$, which deforms the original amplitudes of the superposition and depends on the angles $\varphi_i$ as well as the history of failures and successes in the execution of the repeat-until-success circuit (hence $F_i$ is a random variable). Check the Appendix \ref{sec:rus} for more detailed characterizations.}

Depending on the applications, it might be required or simply desirable to consider parameter settings that give rise to dynamics more closely related to that of classical neural networks. We illustrate one way of achieving this desired behavior for quantum neurons. We relate $\varphi$ to the input signal $\theta$ by the identity $\varphi=\gamma\theta+\pi/4$, where $\gamma=O(1/n)$ is a scaling factor to ensure that $\varphi$ is contained between 0 and $\pi/2$. Furthermore, we restrict the weights and bias values to integer multiples of a finite resolution parameter $\delta$ such that the input signal cannot get arbitrarily close to 0. This puts a lower bound on the value of $\Delta_0\ge\delta/2$. See Appendix \ref{sec:wbsetting} for details.

Combining the above analyses on the minimum $k$ needed for arbitrary error $\epsilon$ (Appendix \ref{subsec:mapq}), the expected runtime as a function of $k$ (Appendix \ref{sec:rus}) and the additional parameter restrictions (Appendix \ref{sec:wbsetting}), we have the following performance guarantee for a quantum neuron. For a detailed proof see Appendix \ref{app:proof}.

\begin{theorem}\label{thm:rustime}
For a quantum neuron with input angle $\varphi$ and $k$ iterations of repeat-until-success circuit, the expected runtime for preparing an output qubit in state $R_y(2q^{\circ k}(\varphi))|0\rangle$ 
such that $|q^{\circ k}(\varphi)-g(\varphi)|\le\epsilon$, where $g(x)=0$ if $x<\pi/4$ and $=\pi/2$ otherwise, is
\begin{equation}
O\left(
({n}/{\delta})^{2.075}(1/\epsilon)^{3.15}
\right)
\end{equation}
where $n$ is the number of input neurons and $\delta$ is a resolution parameter associated with the setting of weights and bias.
\end{theorem}

The quantum neuron we propose can be used as building block for a wide variety of interesting quantum neural network models. We consider two important applications, the first being \emph{feedforward networks}. The template of this kind of network may vary, from shallow ones demonstrated in this article to constructions similar to modern deep learning algorithms.
The second application is the \emph{Hopfield network}, which is a recurrent neural network that exhibits dynamical properties typical of associative memories and attractors. Being able to capture these properties is considered one of the fundamental requirements of a genuine model of quantum neural networks \cite{Schuld2014TheNetwork}.
The goal of our study is two-fold: On one hand we provide rigorous connection between our construction and classical neural networks, and on the other hand we obtain numerical evidence that the networks of quantum neurons can learn from a superposition of training data.

$\quad$\\
\noindent{\bf Feedforward neural network}
\label{sec:ffnn}

Feedforward neural networks have been shown, both theoretically \cite{Auer2008APerceptrons} and empirically \cite{Hinton2006ReducingNetworks}, to capture non-trivial patterns in data. Here we arrange multiple copies of our quantum neuron to reproduce and generalize the behavior of traditional feedforward neural networks. Due to the coherent nature of our construction, we are able to process training inputs in superposition, which may enable one to handle larger training sets than what is typically tractable on classical computers. 
Moreover, the ancilla qubits necessary for the RUS circuit can be reused for all neuron updates, and therefore our construction requires a single qubit for each extra neuron.

Consider a classical feedforward neural network that serves as a binary function $f:\{-1,1\}^n\mapsto\{-1,1\}^m$ that takes an input of $n$ bits and returns a binary string of $m$ bits. The input ${\bf x}\in\{-1,1\}^n$ is stored in the input layer of $n$ neurons. Then the states of the input neurons are passed on to a hidden layer of neurons. The $i$-th hidden neuron will linearly combine the values of the input layer, forming a biased weighted input signal $\theta_i=\sum_jw_{ij}x_j+b_i$ and the final state of the hidden neuron is computed by feeding the input through a step function $\sigma(\theta_i)$ which evaluates to $+1$ if $\theta_i>0$ and $-1$ otherwise.
The values of the hidden neurons may be passed to yet another layer of hidden neurons in the same fashion, until the final layer is reached. This final layer consists of $m$ neurons that store the outputs of the neuron network. Each two adjacent layers of neurons are commonly connected as a complete bipartite graph. The corresponding neural network can be seen as a classifier. For example the input ${\bf x}$ may represent a black-an-white figure with $n$ pixels and the output its classification according to $m$ distinct categories ($+1$ and $-1$ meaning ``belonging'' and ``not belonging'' to each specific category).

Denote $\sigma(\bf y)$ as the vector obtained by applying $\sigma$ element-wise to the vector $\bf y$. 
Consider a multi-layer perceptron with input layer {being in state} ${\bf z}^{(0)}={\bf x}$ and $\ell$ hidden layers in states ${\bf z}^{(1)},\cdots,{\bf z}^{(\ell-1)},{\bf z}^{(\ell)}$ with ${\bf z}^{(\ell)}$ being the state of the output layer. Let ${\bf W}^{(i)}$ be the weight matrix connecting the $(i-1)$-th layer to the $i$-th and ${\bf b}^{(i)}$ be the bias on the $i$-th layer, $i=1,\cdots,\ell$. Then the neural network propagates the information according to the relationship 
\begin{equation}\label{eq:prop}
{\bf z}^{(i)}=\sigma({\bf W}^{(i)}{\bf z}^{(i-1)}+{\bf b}^{(i)}),\quad i=1,\cdots,\ell,
\end{equation}
and the overall function is denoted as ${\bf z}^{(\ell)}=f(\bf x)$.
For a given set of training data $\{{\bf x}_1, {\bf x}_2, \cdots, {\bf x}_T\}$ with the corresponding outputs ${\bf y}_1$, ${\bf y}_2$, $\cdots$, ${\bf y}_T$, the goal of training is to minimize the loss function over the training data:
\begin{equation}\label{eq:obj}
\min_{{\bf W},{\bf b}}\sum_{j=1}^T\left\|{\bf y}_j-f({\bf x}_j)\right\|_2^2.
\end{equation}
Here we have ignored additional terms that may arise in practice such as regularization. Because the objective function as well as the parameters, \emph{i.e.\ }weights and biases, are evaluated classically, these additional terms can be easily added as part of the classical computing for the objective function.

\begin{figure}
\includegraphics[scale=1]{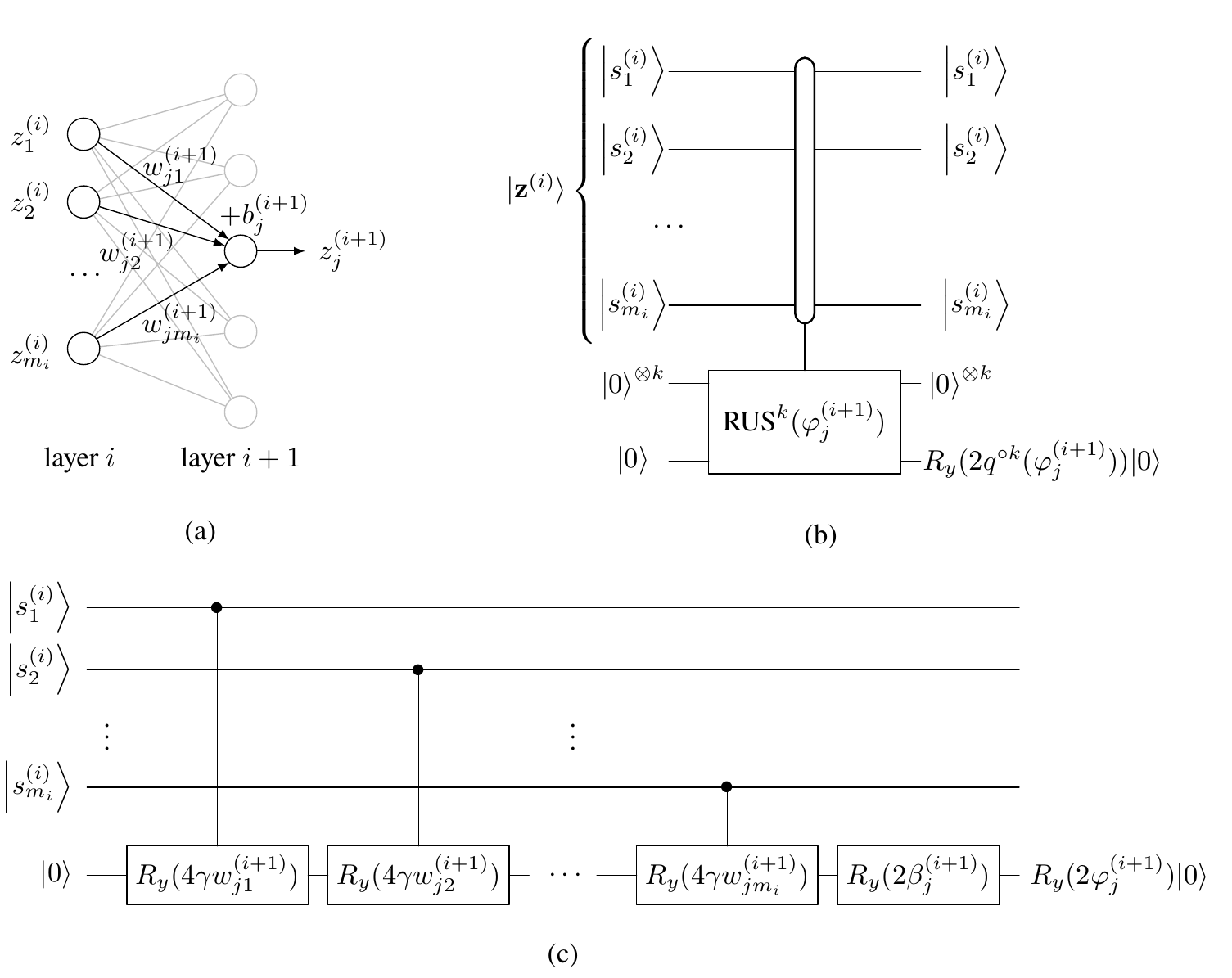}
\caption{Contruction of feedforward network of quantum neurons. \textbf{(a)}
{\bf Propagating the neural state} of the previous layer to a neuron in the current layer of the classical neural network.
\textbf{(b)}
{\bf Quantum circuit realization of the quantum neuron propagation}. Here the bottom qubit corresponds to neuron $z_j^{(i+1)}$. The $k$ ancilla qubits can be recycled for the next neuron propagation. The block control for the $\text{RUS}^k$ operation represents controlled rotations by angle $\varphi_j^{(i+1)}$, as shown in detail in subfigure (c) as well as Figure \ref{fig:rus_twice}.
\textbf{(c)}
{\bf Circuit for applying input rotation} by the angle $\varphi_j^{(i+1)}=\gamma\theta_j^{(i+1)}+\pi/4$, where $\theta_j^{(i+1)}=\sum_{p=1}^{m_i}w_{jp}^{(i+1)}z_p^{(i)}+b_j^{(i+1)}$ is the input signal, $\gamma$ is a scaling factor to ensure that $\varphi_j^{(i+1)}\in[0,\pi/2]$ and $\beta_j^{(i+1)}=\pi/4+\gamma(b_j^{(i+1)}-\sum_{p=1}^{m_i}w_{jp}^{(i+1)})$ is an angular shift (see Appendix \ref{sec:wbsetting} for details). 
}
\label{fig:qffnn}
\end{figure}

In our quantum setting, we introduce one qubit for each neuron in a classical feedforward neural network. We also introduce $k$ ancilla qubits for the RUS circuits. As shown in Figure \ref{fig:qffnn}, the propagation from layer $i$ to each individual neurons in $i+1$ is realized by a $k$ iterations of RUS circuits where the state of the qubits corresponding to the previous layer serve as the control register for determining the input angle $\varphi_j^{(i+1)}$ to the $j$-th qubit in the layer $i+1$. The RUS iterations realizes the threshold dynamics similar to the activation function in the case of classical neural networks. Broadly speaking one could consider our quantum feedforward network a quantized version of a classical neural network where all of the neurons in the hidden layers of the classical network are replaced by quantum neurons. The properties of RUS circuits allow us to approximate Equation \ref{eq:prop} with $\sigma$ being a sigmoid function such as $q^{\circ k}(\varphi)$.

One could use this quantum feedforward neural network for recovering classical feedforward neural networks. The precise connection is stated in the following theorem. Its proof is presented in Appendix \ref{app:ffnn}. Here we say that a quantum algorithm \emph{simulates} a classical feedforward neural network if and only if it produces states $|{\bf z}^{(i)}\rangle$, $i=0,\cdots,\ell$, from which one could efficiently obtain the corresponding state ${\bf z}^{(i)}\in\{-1,1\}^{m_i}$ of each layer $i$ of the classical neural network\footnote{Note that we are not only concerned with replicating the input-output relationship of a neural network, but rather the entire process of neural network propagation. The former notion would correspond to an \emph{emulation} of a neural network rather than simulation.}.

\begin{theorem}\label{thm:cffnn}
There is a quantum algorithm which simulates, with success probability at least $1-\eta$ and error at most $\epsilon$, an $\ell$-layer classical deep feedforward neural network with layer size at most $n$, step function activation and weights/bias setting described in Equation~(\ref{eq:wb}), in time 
\begin{equation}\label{eq:time_complexity}
O\left(\frac{n^{3.075}\ell}{\delta^{2.075}\epsilon^{3.15}}\log\left(\frac{\ell}{\nu}\right)\right).
\end{equation}
The total number of qubits needed for the simulation is
\begin{equation}
O\left(n\ell+\log\frac{n}{\delta\epsilon^{1.52}}\right).
\end{equation}
\end{theorem}

In the setting of the above theorem, we assume that the input state is a computational basis state $|{\bf z}^{(0)}\rangle$ that corresponds to some classical input. However, in general one could imagine having a superposition of training data as inputs (refer to Equation \ref{eq:obj} for notation definition):
\begin{equation}\label{eq:super_input}
\frac{1}{\sqrt{T}}\sum_{j=1}^T \ket{{\bf x}_j} \ket{{\bf y}_j}
\end{equation}
where the register holding $\ket{{\bf x}_j}$ is the input layer of the quantum neural network and we introduce an ancilla qubit for holding the state $\ket{{\bf y}_j}$, which is the correct output for each of the training data ${\bf x}_j$. We do not make assumption concerning the method to efficiently generate the superposition of the states $\ket{{\bf x}_j} \ket{{\bf y}_j}$.
It can be obtained from classical training examples by using a construction such as the QRAM \cite{Giovannetti2008QuantumMemory}, or may be the output of a quantum circuit, or simply provided by the owner of quantized databases.
The information contained in the state described by expression~(\ref{eq:super_input}) is then propagated through the quantum neural network. Depending on the specific task considered, there are two ways to quantify the accuracy of training. The first one is by averaging over the pairwise measurement of $\langle ZZ\rangle$ on the output layer qubits in state $|\tilde{\bf z}^{(\ell)}\rangle$ and the ancilla qubits with the state $|{\bf y}_j\rangle$. 
More precisely, suppose ${\bf y}_j$ has length $m$. Let $a_1$ through $a_m$ denote qubits in $|\tilde{\bf z}^{(\ell)}\rangle$ and $b_1$ through $b_m$ denote qubits in $|{\bf y}_j\rangle$. The training accuracy can be characterized by
\begin{equation}\label{eq:bar_ZZ}
\overline{\langle ZZ\rangle}=\frac{1}{m}\sum_{j=1}^m\langle Z_{a_j}Z_{b_j}\rangle.
\end{equation}
The averaged expectation value $\overline{\langle ZZ\rangle}$ ranges between $-1$ and $1$, with $\overline{\langle ZZ\rangle}=1$ signifying perfect training. Hence we formulate our training problem as finding the assignment of weights and biases satisfying Equation~(\ref{eq:wb}) such that the $\overline{\langle ZZ\rangle}$ value is maximized. The second way to quantify the training accuracy is by computing the product of the pairwise $\langle ZZ\rangle$ values:
\begin{equation}\label{eq:bar_ZZ2}
\overline{\langle ZZ\rangle}=\prod_{j=1}^m\langle Z_{a_j}Z_{b_j}\rangle.
\end{equation}
In this case $\overline{\langle ZZ\rangle}=1$ still signifies perfect training. However, such training objective puts more stringent requirement on the classifier being able to produce high accuracy on all of the qubits. 

We use two examples to illustrate the design of a quantum neural network that implements a binary classifier. The first example is the \emph{XOR function}, where the goal is to train a neural network as a binary classifier to compute the function $XOR(x_1,x_2)=x_1\oplus x_2$ for any given input $(x_1,x_2)\in\{0,1\}^2$. The XOR problem is important because it was one of the first problems identified as an example of the limitations of the common artificial neuron construction, since a single neuron, or a single layer of neurons, cannot capture the linearly inseparable XOR function \cite{minsky_papert_1969}. However, by constructing a multi-layer neural network such as the 2-2-1 network (Figure \ref{fig:xor_ffnn}a), one could train the network to learn the XOR function. During the training we iteratively vary the weights and biases to maximize $\overline{\langle ZZ\rangle}$ between the output qubit and the training qubit, and test the current parameter setting by initializing the input and training registers at individual $|{\bf x}_j\rangle|{\bf y}_j\rangle$ states (Equation \ref{eq:super_input}) and evaluate $\overline{\langle ZZ\rangle}$. We then define the accuracy values $(\overline{\langle ZZ\rangle}+1)/2$ obtained from training and testing as the \emph{training accuracy} and \emph{testing accuracy} respectively. 

Unlike classical neural networks, whose training is typically based on gradient descent and its variants, in our case gradient-based optimization would be challenging due to the stochastic nature of the RUS circuits and therefore the objective function \cite{Guerreschi2017PracticalAlgorithms}. Instead we choose to use the Nelder-Mead algorithm \cite{Nelder1965,Kelley1999,Nocedal2000}, which is a gradient-free local optimization method, for training the feedforward networks of quantum neurons.

Numerical results (Figure \ref{fig:xor_ffnn}b) show that the network can be trained to almost perfect training and testing accuracy. Note that the training process is based on only copies of \emph{a single state} of the form in \eqref{eq:super_input}, while the testing is performed on \emph{each input state $|{\bf x}_j\rangle|{\bf y}_j\rangle$ separately}. This demonstrates a form of learning which is fundamentally different from the classical neural networks, where training data are fed sequentially into the training as mini-batches. Our numerical results provide evidence that it suffices to train the quantum neural network on a single state which is a superposition of training data.

\begin{figure}
\hspace*{-0.3in}
\includegraphics[scale=0.8]{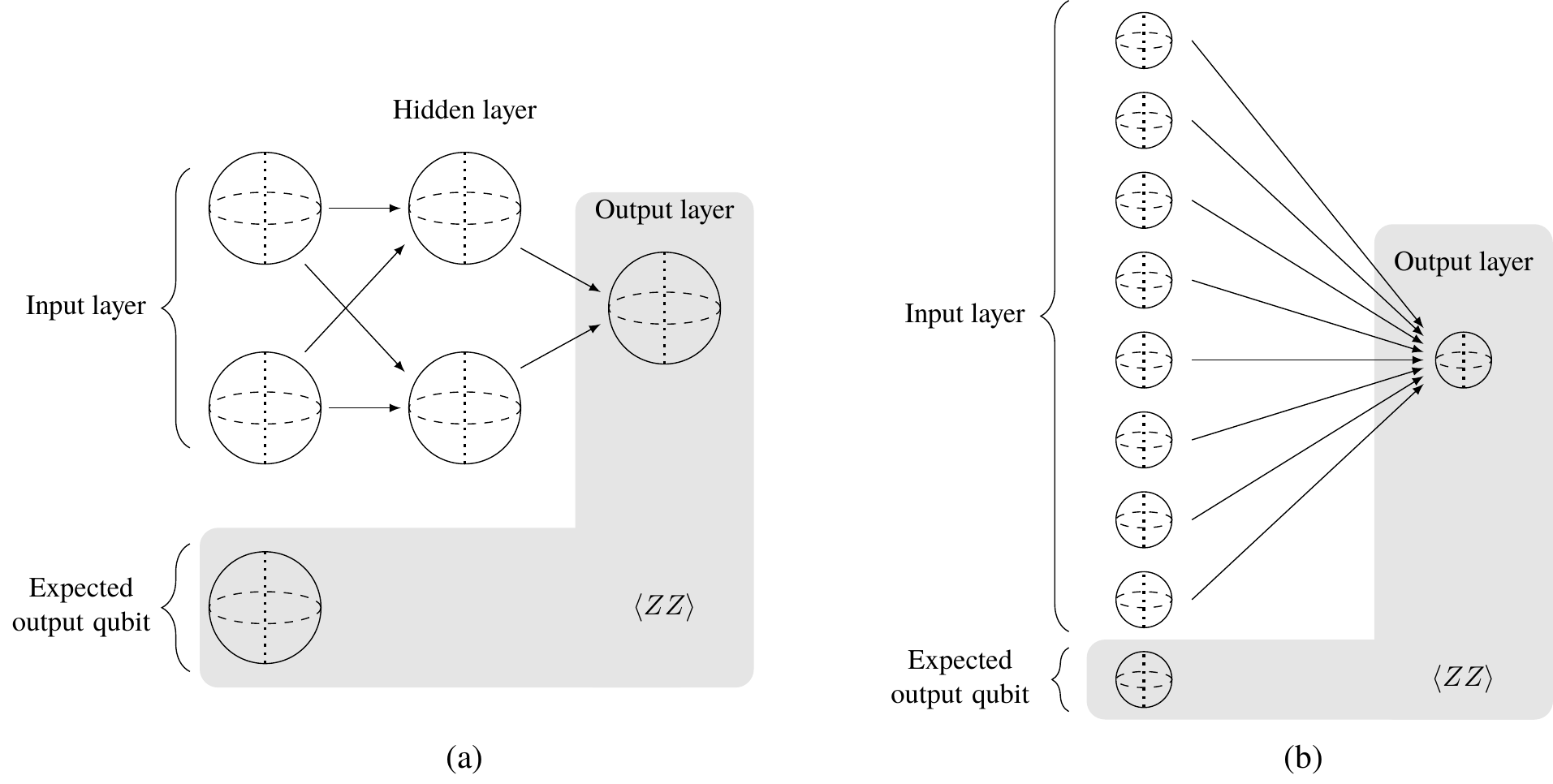}
\caption{Quantum neural network able to learn the {\bf (a) XOR function} and {\bf (b) 8-bit Parity function}. Here each sphere represents a qubit. Unlike Figure \ref{fig:basic_setting}b, for simplicity we omit the ancilla qubit in the arrow representations of the propagation of quantum state from one layer to the next. During training, the input layer is initialized in a superposition of the training examples, and the state of the hidden and output layer obtained through the RUS method. Finally, one measures the observable $\langle ZZ\rangle$ between the output qubit and the expected output qubit.}
\label{fig:network-designs}
\end{figure}

\begin{figure}
\hspace*{-0.3in}
\includegraphics[scale=0.7]{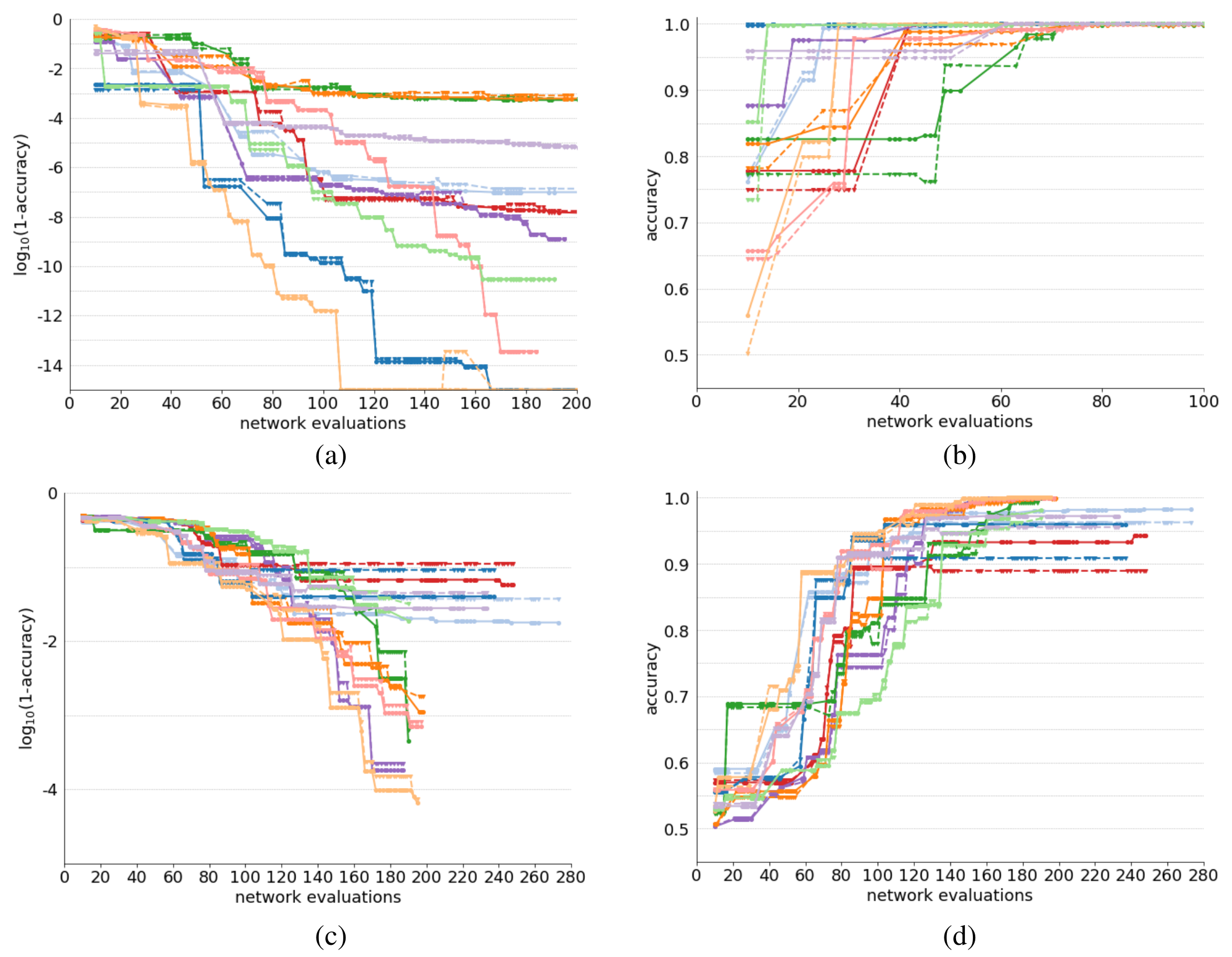}
\caption{Training feedforward networks of quantum neurons. {\bf (a) Results for optimizing the parameters of the XOR network} (Figure \ref{fig:network-designs}a) using Nelder-Mead algortihm. Here ``accuracy'' is defined as $(\overline{\langle ZZ\rangle}+1)/2$ where $\overline{\langle ZZ\rangle}$ is defined in Equation \ref{eq:bar_ZZ}. We train on a superposition of training data but test on individual computational basis states $|x_1x_2\rangle|XOR(x_1,x_2)\rangle$. The solid lines represent the training accuracy while the dashed lines represent the testing accuracy. Different colors represent optimization runs with different initial guesses. {\bf (b) A linear scale plot of accuracy versus the number of iterations} (same data as subplot (a)) to highlight the difference between training and testing accuracy in the initial iterations. {\bf (c) Results for optimizing the parameter of the 8-1 parity network} (Figure \ref{fig:network-designs}b). We train on the state which is an even superposition of $|x_1\cdots x_8\rangle|Parity(x_1,\cdots,x_8)\rangle$ and compute testing accuracy in the same way as in subplot (a). {\bf (d) A linear scale plot of accuracy versus the number of iterations} (same data as subplot (c)).}
\label{fig:xor_ffnn}
\end{figure}

To add to the numerical evidence we have used a second example which is an \emph{8-1 parity network}. A schematic of the network is shown in Figure \ref{fig:network-designs}a. The goal is to train the network to learn an 8-bit parity function $Parity(x_1,\cdots,x_8)=x_1\oplus \cdots \oplus x_8$. The numerical results are shown in Figure~\ref{fig:network-designs}b, which shows that for some initial guesses of the weights, the network can be trained to perfect accuracy. Unlike the XOR network which has a hidden layer, here the parity network does not have any hidden layer while it is still able to learn the parity function. This is because our training does \emph{not} impose restriction on the range of weights and biases, unlike the setting of Theorem~\ref{thm:cffnn}. In other words, for both numerical examples $\gamma=1$ (see Figure \ref{fig:qffnn}c for definition of $\gamma$) instead of $O(1/n)$ as discussed in Appendix~\ref{sec:wbsetting}. This enables the training to take advantage of the periodicity of the function $q(\theta)$, since unrestricted weights and biases may put $\theta$ outside the interval $[0,\pi/2]$, where the function $q(\theta)$ is not strictly monotonic. In fact, such periodicity allows us to simplify the XOR network by removing the hidden layer without affecting the overall training / testing accuracy. 

$\quad$\\
\noindent{\bf Hopfield network}

\begin{figure}
\begin{center}
\includegraphics[scale=1]{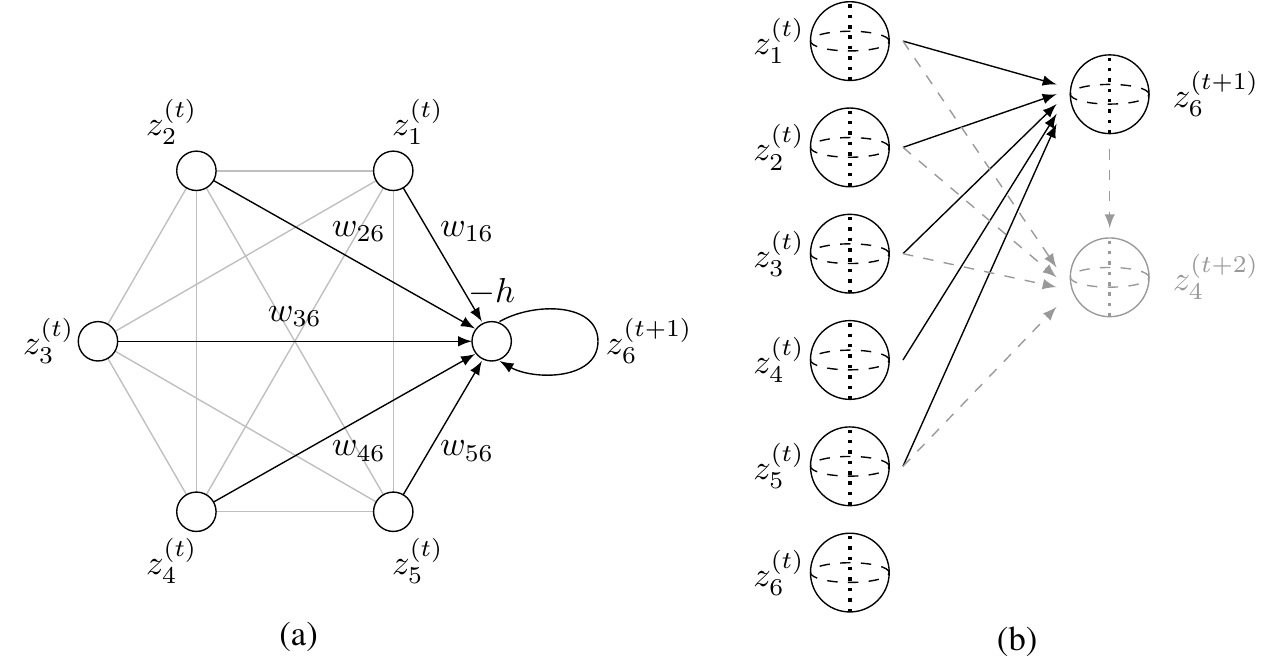}
\end{center}
\caption{Schematics of a Hopfield network update step. {\bf (a) Classical Hopfield network update}. Here at step $t$, the neuron $6$ is chosen for update. The new state $z_6^{(t+1)}=\sigma(w_{16}z_1^{(t)}+\cdots+w_{56}z_5^{(t)}-h)$ where $\sigma$ is a step function such that $\sigma(x)=1$ if $x>0$ and $-1$ if $x<0$. Note the minus sign before $h$ because in this case $h$ is the \emph{threshold} rather than the bias, unlike feedforward networks. {\bf (b) Quantum Hopfield network update}. Here the solid arrows indicate propagation with RUS circuits in the same fashion as Figure \ref{fig:qffnn} during the update from time $t$ to $t+1$ where qubit $6$ is chosen, except for the sign reversal in the handling of thresholds. The dashed arrows indicate the scheme for updating from $t+1$ to $t+2$ where qubit $4$ is chosen.}
\label{fig:hn}
\end{figure}

In the classical setting, a Hopfield network starts with an initial state $(z_1,z_2,\cdots,z_n)$ which can be considered as the input state. The state of each neuron $z_i\in\{-1,1\}$ is a classical bit.
Then the network undergoes a sequence of updates. During each update, a random neuron $j$ is selected and its new state is assigned to be $1$ if $\sum_{i\neq j}w_iz_i>h_j$ for some threshold $h_j$ associated with the $j$-th neuron and the new state is assigned to be $-1$ otherwise. As the updates proceed, the state of the network converges to a state that is a local minimum of the energy function $E=-\frac{1}{2}\sum_{i,j}w_{ij}z_iz_j+\sum_ih_iz_i$. Such local minima are \emph{attractors} of the network and any input state that is reasonably close to a local minimum will converge to the minimum after sufficiently many updates. It is such attractor dynamics that gives a Hopfield network the property of an \emph{associative memory}. 

In the quantum setting, we introduce $n$ qubits, one for each neuron in the classical Hopfield net. Assume the $n$ qubits are in a computational basis state corresponding to the state of the classical Hopfield net before the update. 
To realize the threshold dynamics of the update step, we introduce $k$ ancilla qubits for the iterative repeat-until-success circuits (Figure \ref{fig:rus_twice}b). For the $j$-th update, suppose neuron $i_j\in[n]$ is chosen. Then we use the joint state of the remaining qubits $[n]\backslash\{i_j\}$ as the input state $|\varphi\rangle$ for the RUS circuit (Figure \ref{fig:rus_twice}) which produces an output qubit that is close to the state $|0\rangle$ if the total weight $\sum_{j\neq i_1}w_{ji_1}z_j<h_{i_1}$ and $|1\rangle$ otherwise. The detailed circuit for realizing such transformation is similar to the ones in Figure \ref{fig:qffnn}, with minor modifications of the bias replaced with minus the threshold (Figure \ref{fig:hn}a). We call the output qubit of the RUS circuit the new neuron $i_j$ and use its state for the next updates. In general, let $b_i^{(j-1)} \in\{0,1\}$ be the state of neuron $i\in[n]$ at the $(j-1)$-st update: $b_i^{(j-1)}=0$ iff $z_i^{(j-1)}=-1$ and $b_i^{(j-1)}=1$ iff $z_i^{(j-1)}=1$. Then suppose neuron $i_j$ is chosen for the $j$-th update. We use the joint state of qubits corresponding to the latest state of neurons $n\backslash\{i_j\}$ as the input state for the RUS circuits, producing a new qubit in state $|b_{i_j}^{(j)}\rangle$ (Figure \ref{fig:hn}b). After a sequence of $t$ updates, we have a state of the form
\begin{equation}\label{eq:hopstate}
|b_1^{(0)}\cdots b_n^{(0)}b_{i_1}^{(1)}b_{i_2}^{(2)}\cdots b_{i_t}^{(t)}\rangle
\end{equation}
where $b_{i_k}^{(k)}$ is the state of neuron $i_k$. The state above records a history of how the state has evolved over the updates. If at the $(t+1)$-st update, the neuron $i_{t+1}$ is chosen, then we introduce a new qubit in $|0\rangle$ state and apply $k$-iteration RUS (Figure \ref{fig:rus_twice}b) using the state of the set of qubits representing the latest state of the remaining neurons $[n]\backslash\{i_{t+1}\}$ as the control state $|\varphi\rangle$. 

The behaviour of classical Hopfield network can also be easily recovered from this quantum model. For simulating classical Hopfield network, we do not even need the number of qubits to grow linearly as the number of updates. Suppose at some iteration $k$, the neuron $i_k$ is chosen for update. We apply the RUS iterations as before to prepare an output state which is close to either $|0\rangle$ or $|1\rangle$ depending on the states of the neurons (qubits) $[n]\backslash\{i_k\}$. We can simply measure the output qubit, and repeat the RUS propagation and output qubit measurement several times. A majority vote on the measurement outcomes should give us the correct state with high probability. We can then prepare the qubit $i_k$ in the $n$ qubits for the Hopfield network in the same state as the output qubit.

A precise statement regarding the connection between the Hopfield network of quantum neurons and the classical Hopfield network is the following theorem below. Its proof is stated in Appendix \ref{app:hopfield}. Here we say a network of quantum neuron \emph{simulates} $t$ updates of a Hopfield network of $n$ neurons if and only if for a sequence of intermediate states ${\bf z}^{(0)}$, ${\bf z}^{(1)}$, $\cdots$, ${\bf z}^{(t)}$ of the Hopfield network, the quantum system  also goes through a sequence of states  $\ket{{\bf z}^{(0)}}$, $\ket{{\bf z}^{(1)}}$, $\cdots$, $\ket{{\bf z}^{(t)}}$ with each $\ket{{\bf z}^{(i)}}$ such that using it one could efficiently compute ${\bf z}^{(i)}$.

\begin{theorem}
There is a quantum algorithm that simulates $t$ updates of an $n$-neuron Hopfield network with weights and biases satisfying Equation \ref{eq:wb}, up to error $\epsilon$ and success probability at least $1-\nu$, in expected runtime
\begin{equation}
O\left(\frac{n^{2.075}t}{\delta^{2.075}\epsilon^{3.15}}\log\left(\frac{t}{\nu}\right)\right).
\end{equation}
The total number of qubits needed for the simulation is
\begin{equation}
O\left(n+\log\frac{n}{\delta\epsilon^{1.52}}\right).
\end{equation}
\end{theorem}

In the classical case the time and qubit costs are $O(nt)$ and $n$ respectively. Hence our Hopfield network of quantum neurons simulates the classical Hopfield work with a roughly linear (in $n$) overhead in time and logarithmic (in $n$) overhead in memory. However, beyond the classical setting, our Hopfield network of quantum neurons is also capable of behaviours that are unique to quantum systems.

\begin{figure}
\includegraphics[scale=1]{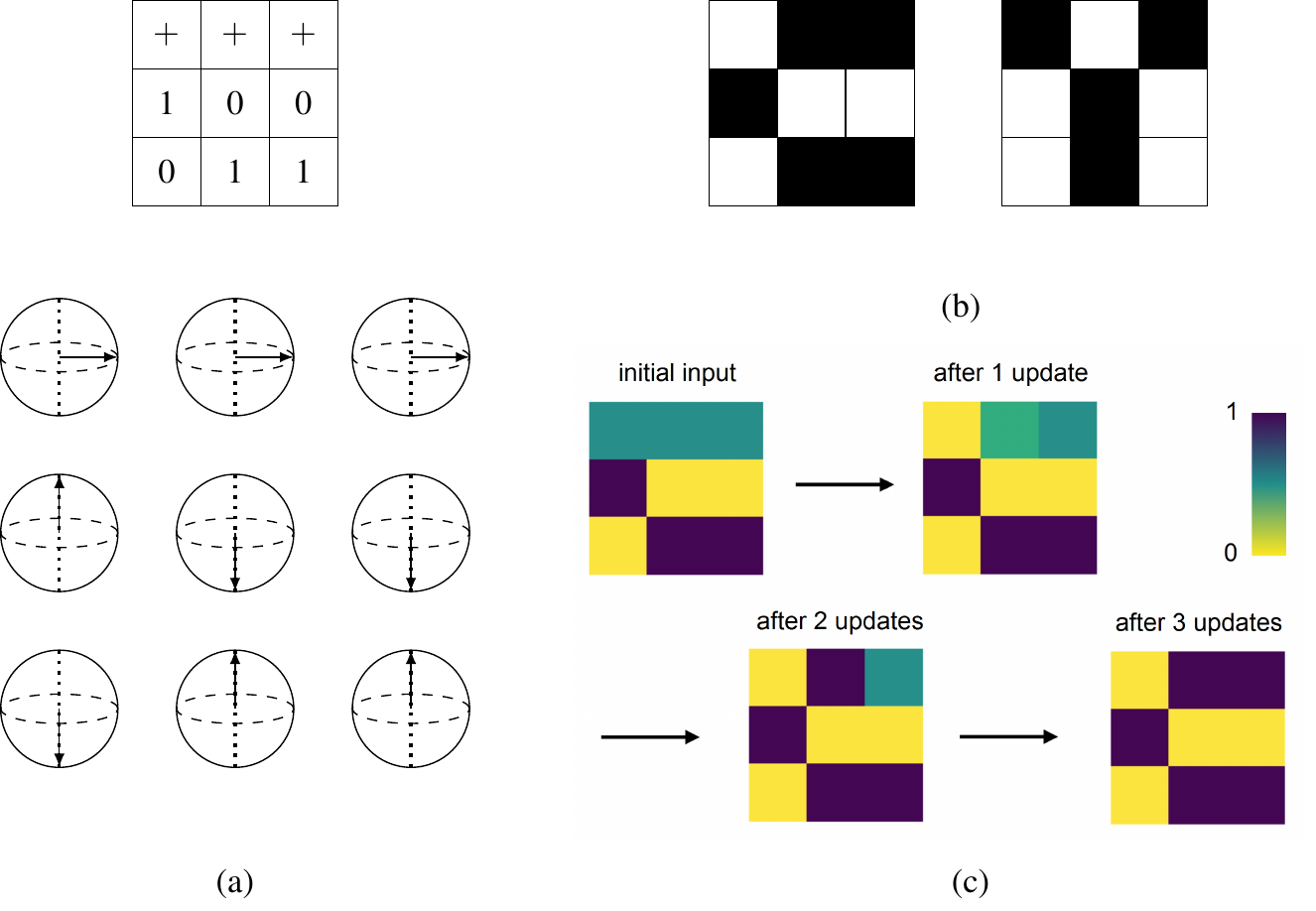}
\caption{Numerical example for a Hopfield network of quantum neurons. {\bf (a) Initial state of the Hopfield network}. Here we corrupt the top three qubits to be in $|+\rangle=\frac{1}{\sqrt{2}}(|0\rangle+|1\rangle)$ state, while the remaining qubits are in the state of the `C' attractor (see subfigure b). {\bf (b) Attractors of the network}. We train the network weights such that the network has the two patterns (`C' and `Y') as its attractors. {\bf (c) Updating the network} to demonstrate the attractor mechanism. We use a color scale from yellow to purple to indicate the marginal probability of each cell (qubit) being in state $|0\rangle$ or $|1\rangle$, with the former represented as entirely white and the latter represented as entirely black. Here the initial state is closest to the `C' attractor and therefore converges to `C'. The three updates are performed on the top row from left to right.}
\label{fig:hop_ex}
\end{figure}

Consider a Hopfield network of quantum neurons arranged as a $3\times 3$ grid of binary pixels (Figure \ref{fig:hop_ex}a). Using Hebbian rules, the network weights are determined such that the two states with `C' and `Y' patterns are the attractor states (Figure \ref{fig:hop_ex}b). 
The network is initialized in the state of corrupted `C' pattern where the top 3 qubits are prepared in the $|+\rangle$ state (Figure \ref{fig:hop_ex}a). Choosing the three corrupted qubits from left to right as the updated qubit, with each update step carried out as in Figure \ref{fig:hn}, the attractor character of the network can be recovered by inspecting the marginal probabilities of each qubit being in state $|0\rangle$ or $|1\rangle$ (Figure \ref{fig:hop_ex}c). The dissipative nature of the attractor dynamics in the Hopfield network of quantum neurons comes from the method for gleaning the current state of the network from the global state in \eqref{eq:hopstate}, which restricts to the subsystem of qubits that correspond to the latest state of each neuron in the Hopfield network. 

$\quad$\\
\noindent{\bf Discussion}

There are a wide variety of proposals for quantum neural networks and in Schuld et al.\ \cite{Schuld2014TheNetwork} the authors survey the literature and argue that a reasonable construction of quantum neural network should satisfy the following requirements:
\begin{enumerate}
\item The initial state of the quantum system encodes any binary string of length $N$;
\item The QNN reflects one or more basic neural computing mechanisms;
\item The evolution is based on quantum effects, such as superposition, entanglement and interference, and it is fully consistent with quantum theory.
\end{enumerate}
Our proposal satisfies the first criterion in an intuitive way, since we map each neuron to a qubit. The set of $N$-qubit computational basis states can naturally represent the set of $N$-bit strings. From the discussion on the basic construction of our quantum neuron (Figure \ref{fig:basic_setting}), our realization of threshold dynamics using RUS circuits reflects the integrate-and-fire mechanism of neural computation. Furthermore, the two examples considered in the paper have demonstrated the application of these neural computing mechanisms to learn Boolean functions and evolve as an associative memory. Therefore our construction also satisfies the second criterion. Finally, we have numerical evidence showing that feedforward networks of quantum neurons can learn from a superposition of training data (Figure \ref{fig:xor_ffnn}) and Hopfield networks of quantum neurons can recover attractors from a quantum superposition of corrupted attractor states (Figure \ref{fig:hop_ex}), where quantum superposition, entanglement and coherence play crucial roles.

Although we have shown feedforward network and Hopfield network of quantum neurons as two examples, the variety of possible uses of the quantum neuron by no means are restricted to these two. Here we discuss a few additional aspects.

\begin{figure}
\includegraphics[scale=0.9]{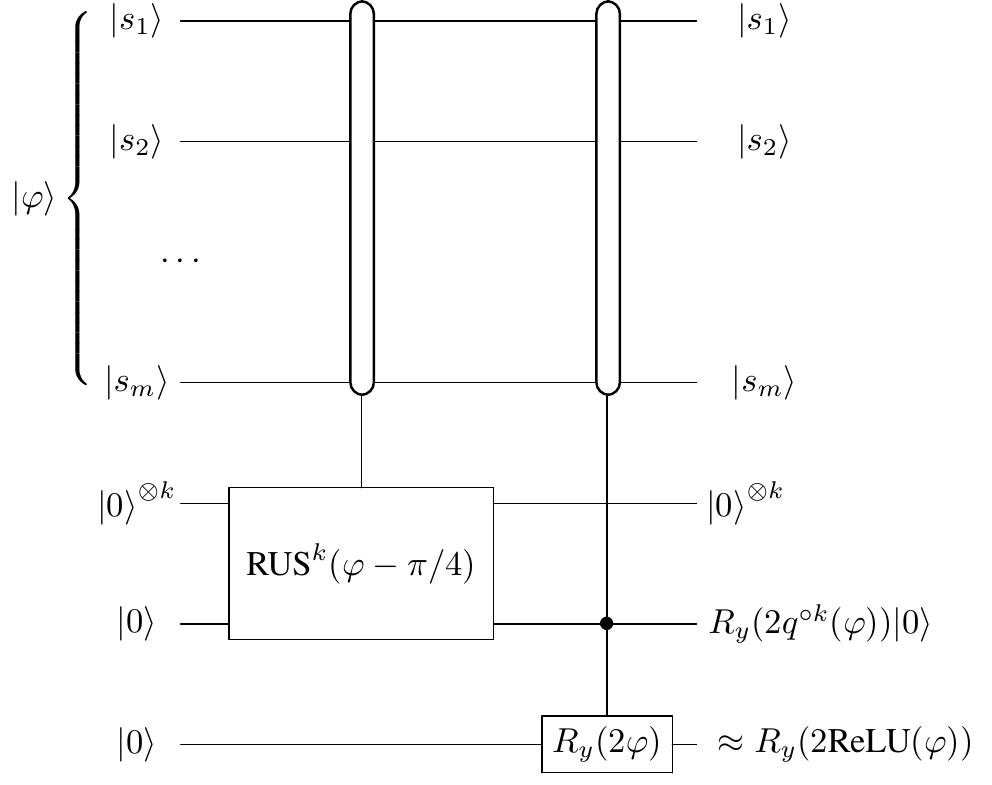}
\caption{Circuit for approximating ReLU activation function.}
\label{fig:relu}
\end{figure}

\begin{enumerate}
\item
\noindent{\textbf{\textit{Network architecture}}}. One could use quantum neurons to construct \emph{quantum autoencoders} \cite{Romero2016QuantumData,Wan2016QuantumNetworks}. The idea is to use the construction of deep feedforward network, but with input and output layers consisting of equal number of qubits {and hidden layers with fewer units}. In training the network, instead of preparing the state in \eqref{eq:super_input}, one prepares the input layer of qubits in whichever state that is to be compressed and measure the quality of the autoencoder by making correlated $\langle ZZ\rangle$ measurements between pairs of qubits in the input and output layers (instead of the training register and output layers as is the case for feedforward network). As another example, we can realize \emph{convolutional networks of quantum neurons} by restricting the connectivity and imposing translational invariance of the weights in the network. With more thoughts and specific applications in mind, one may well be able to find more architectures of quantum neural networks, since the versatility of our quantum neuron construction enables a wide variety of quantum neural networks to be devised. 

\item 
\noindent{\textbf{\textit{Activation function.}}} Another extension of the quantum neuron construction is to approximately realize Rectified Linear Unit (ReLU) activation using the circuit in Figure \ref{fig:relu}. Here we use the circuit in Figure \ref{fig:qffnn}b for producing a qubit that is close to $|1\rangle$ only when $\varphi>0$. This bit is then used as a control qubit, together with the input register encoding $|\varphi\rangle$, for realizing a controlled rotation by angle $\varphi$ on the output qubit (bottom qubit in Figure \ref{fig:relu}). Generalizing this idea one could in principle realize more general forms of piecewise linear activation functions.

\item
\noindent{\textbf{\textit{Paradigm of machine learning}}}. Broadly speaking the field of machine learning consists of three main paradigms, namely supervised, unsupervised and reinforcement learning. We have numerically demonstrated the potential of using our quantum neuron construction for supervised learning in ways that are not possible on classical computers \emph{i.e.\ }training with superposition of input-output pairs. We have also argued that the quantum neuron construction can be used for building unsupervised learning algorithms such as the autoencoder, though further numerical evidence remains to be gleaned. For reinforcement learning, with the capacity for convolutional network and feedforward network, combined with the possibility of realizing ReLU activation, one can in principle use quantum neurons to construct a quantum neural network analogous to that in \emph{e.g.\ }the deep $Q$-network \cite{Mnih2013PlayingLearning,Arulkumaran2017ALearning} for learning control sequences for video games. Suppose the network takes a quantum state representing the superposition of a stack of greyscale frames from the video game and processes it with convolutional and fully connected layers, with ReLU nonlinearities in between each layer. At the final layer, the network outputs a discrete action, which corresponds to one of the possible control inputs for the game. Given the current state and chosen action, the game (classically) returns a new score, which determines the reward. We then adjust the parameters of the network to maximize the reward.
\end{enumerate}

To conclude, we have presented a simple construction of quantum neuron which can be used as a building block for constructing a wide variety of quantum neural networks. We argue that our proposal satisfies the basic requirements \cite{Schuld2014TheNetwork} for producing ``reasonable'' constructions of quantum neural networks. We rigorously show how our neural network constructions reduce to classical networks in certain settings. Additionally, using numerical examples we have demonstrated that feedforward networks of quantum neurons are able to learn non-trivial functions using a superposition of training data, which is beyond what is possible on a classical computer.  Finally, we envision that the quantum neuron be applied to build quantum algorithms for various paradigms of machine learning, including supervised, unsupervised and reinforcement learning. 

\section*{Acknowledgements}
We thank Jonathan Olson for critical reading of the manuscript. YC and AA-G acknowledge National Science Foundation Grant \# CHE-1655187. AA-G acknowledges support from the Vannevar Bush Faculty Fellowship program sponsored by the Basic Research Office of the Assistant Secretary of Defense for Research and Engineering and funded by the Office of Naval Research through grant N00014-16-1-2008.





\bibliographystyle{unsrt}
\bibliography{quantum_neuron}

\newpage

\begin{widetext}
\section*{SUPPLEMENTARY INFORMATION}

\subsection{Convergence analysis of the nonlinear map \texorpdfstring{$q$}{q}}
\label{subsec:mapq}

Here we analyze in detail the nonlinear map $\varphi_{i+1}=q(\varphi_i)=\text{arctan}\tan^2\varphi_i$ to find the minimum number of RUS iterations needed for yielding a final $\varphi_i$ value that differs from its attractor by at most $\epsilon$.  Suppose $\varphi_0>\pi/4$. Let $\Delta_i=\varphi_i-\pi/4$ and $\Omega_i=\pi/2-\varphi_i$.  Then for $\pi/4<\varphi<3\pi/8$, we have
\begin{equation}
\Delta_{i+1}\ge(\pi/8)^{-1}\text{arctan}\tan^2(\pi/8+\pi/4)\Delta_i=\alpha\Delta_i
\end{equation}
where $\alpha\approx 3.5673$. For $3\pi/8<\varphi_i<\pi/2$, we have
\begin{equation}
\Omega_{i+1}\le(\pi/8)^{-1}(\pi/2-\text{arctan}\tan^2(\pi/8+\pi/4))\Omega_i=\beta\Omega_i
\end{equation}
where  $\beta\approx 0.4327$.  Then the number of RUS iterations needed for $\varphi_i$ to reach $3\pi/8$ is no less than $\log_\alpha(3\pi/8\Delta_0)$.  The number of RUS iterations needed for $\varphi_i$ to be $\epsilon$-close to $\pi/2$, which is the attractor for the case $\varphi_0>\pi/4$ we are considering, is no less than $\log_{1/\beta}(1/\epsilon)$. Hence the total number of iterations 
\begin{equation}\label{eq:k_scale}
k=\left\lceil\log_\alpha\left(\frac{3\pi}{8\Delta_0}\right)\right\rceil+\left\lceil\log_{1/\beta}\left(\frac{1}{\epsilon}\right)\right\rceil
\end{equation}
can ensure that $|\varphi_k-\pi/2|\le\epsilon$. Because of symmetry the same expression for $k$ can be derived for the case $\varphi_0<\pi/4$. 

\subsection{Runtime analysis of RUS circuits}\label{sec:rus}

Consider the RUS circuit in Figure \ref{fig:basic_setting}c.  A simple calculation shows that the probability of measuring $|0\rangle$ is $p(\theta)=\sin^4\theta+\cos^4\theta$ and the probability of measuring $|1\rangle$ is  $1-p(\theta)$.  Hence the expected runtime of the circuit is $1/p(\theta)\le 2$. For $k$ iterations of RUS circuits, there are in total $k$ different levels of recursions, with the bottom  recursion consisting of circuits of Figure \ref{fig:basic_setting}c. Let $t_j$ be the total time spent running $j$ iterations of the RUS circuit (For example in Figure \ref{fig:rus_twice}, $j=2$).  Then we have expectation values 
\begin{equation}
\begin{array}{ccl}
\mathbb{E}(t_1) & = & \displaystyle \frac{1}{p(\theta)}\le 2 \\
\mathbb{E}(t_k) & = & \displaystyle \frac{1}{p(q^{\circ k}(\theta))}\cdot\mathbb{E}(t_{k-1})\cdot 2=\left(\prod_{k'=0}^kp(q^{\circ k'}(\theta))\right)^{-1}\cdot 2^{k+1}\le 2^{2k+1}.
\end{array}
\end{equation}
Now let us consider RUS iterations with the base recursion consisting of the circuit in Figure \ref{fig:rus_super} with the control register being in a general superposition $\sum_{i=1}^T\alpha_i|i\rangle$. For a single iteration (Figure \ref{fig:rus_super}), the state of the system before measure is
\begin{equation}\label{eq:Ir}
\sum_{i=1}^T\alpha_i|i\rangle\otimes\left[\sqrt{p(\varphi_i)}|0\rangle R_y(q(\varphi_i))|0\rangle+\sqrt{1-p(\varphi_i)}|1\rangle R_y(\pi/4)|0\rangle\right].
\end{equation}
The probability of measuring $|0\rangle$ at this stage is $P=\sum_{i=1}^T|\alpha_i|^2p(\varphi_i)$, yielding a state
\begin{equation}
\sum_{i=1}^T\alpha_i\sqrt{\frac{p(\varphi_i)}{P}}|i\rangle|0\rangle R_y(q(\varphi_i))|0\rangle.
\end{equation}
The probability of measuring $|1\rangle$ is $P^\perp=\sum_{i=1}^T|\alpha_i|^2(1-p(\varphi_i))$, yielding a state
\begin{equation}
\sum_{i=1}^T\alpha_i\sqrt{\frac{1-p(\varphi_i)}{P^\perp}}|i\rangle|1\rangle R_y(\pi/4)|0\rangle.
\end{equation}
In general if RUS fails $r-1$ times and succeeds (namely measuring $|0\rangle$) at the $r$-th trial, we have a state
\begin{equation}\label{eq:Ir}
\sum_{i=1}^T\alpha_i\sqrt{\frac{(1-p(\varphi_i))^{r-1}p(\varphi_i)}{P_{r}}}|i\rangle|0\rangle R_y(q(\varphi_i))|0\rangle
\end{equation}
where $P_r=\sum_{i=1}^T|\alpha_i|^2(1-p(\varphi_i))^{r-1}p(\varphi_i)$ is the normalization factor for the state corresponding to success at trial $r$. Accordingly we define $P^\perp_{r}=\sum_{i=1}^T|\alpha_i|^2(1-p(\varphi_i))^{r}$ as the normalization factor for the state produced at failure of the trial $r$.  At the $r$-th trial, for $r>1$ the probability of success is $P_r/P_{r-1}^\perp$ and the probability of failure is $P_r^\perp/P_{r-1}^\perp$.  Hence the expected number of trials needed is 
\begin{equation}\label{eq:Et}
\mathbb{E}(t) = P_1 + \sum_{r=2}^\infty r\frac{P_r}{P_1^\perp}\le 1+\sum_{r=2}^\infty r\left(\frac{1}{2}\right)^{r-2}=7. 
\end{equation}
Here we have used the inequalities $P_r\le\sum_{i=1}^T|\alpha_i|^2(1-p(\varphi_i))(1/2)^{r-2}=P_1^\perp(1/2)^{r-2}$ as well as $1/2\le p(\varphi)\le 1$.  Consider the second round of RUS circuit with the input qubit reset to $|0\rangle$ after the measurement step (circuit (II) in Figure \ref{fig:rus_appendix}). Suppose the subroutine (I) underwent $r$ trials, producing a state in Equation \ref{eq:Ir}. Let $\alpha'_i=\alpha_i\sqrt{(1-p(\varphi_i))^{r-1}p(\varphi_i)/P_r}$. Then the probability analysis for circuit (II) is analogous to that for circuit (I). The state after $s$ trials with the last trial being the only success can be written as\footnote{In principle we should write $p(-\varphi_i)$ and $q(-\varphi_i)$ instead of $p(\varphi_i)$ and $q(\varphi_i)$. However, since $p(\varphi)=p(-\varphi)$ and $q(\varphi)=q(-\varphi)$ for any $\varphi$ we use $\varphi_i$ instead.}
\begin{equation}
\sum_{i=1}^T\alpha_i'\sqrt{\frac{(1-p(\varphi_i))^{s-1}p(\varphi_i)}{P'_s}}|i\rangle\left[\sqrt{p(q(\varphi_i))}|0\rangle R_y(q(q(\varphi_i)))|0\rangle+\sqrt{1-p(q(\varphi_i))}|1\rangle R_y(\pi/4)|0\rangle\right]
\end{equation}
where $P'_s=\sum_{i=1}^T|\alpha'_i|^2(1-p(\varphi_i))^{s-1}p(\varphi_i)$. Define $P_s^{'\perp}=\sum_{i=1}^T|\alpha'_i|^2(1-p(\varphi_i))^s$. Then the expected number of trials needed for (II) is
\begin{equation}\label{eq:EtII}
\mathbb{E}(t_\text{II})=P'_1+\sum_{s=2}^\infty s\frac{P'_s}{P_s^{'\perp}}\le 7
\end{equation}
by the same arguments leading up to Equation \ref{eq:Et}. The reason for the similarity between Equation \ref{eq:Et} and \ref{eq:EtII} is that the arguments leading up to  Equation \ref{eq:Et} is independent of $\{\alpha_i\}$. Therefore no matter whether it is $\{\alpha_i\}$ or $\{\alpha'_i\}$, essentially every RUS iteration of every level has expected number of trials bounded from above by 7.  The expected number of bottom recursion trials in $k$ iterations of RUS can then be bounded from above as
\begin{equation}\label{eq:Etk}
\mathbb{E}(t_k)\le 7\cdot\mathbb{E}(t_{k-1})\cdot 2 \le\cdots\le 14^{k-1}\mathbb{E}(t_1)\le \frac{1}{2}\cdot 14^k.
\end{equation}

\begin{figure}
\begin{center}
\includegraphics[scale=1]{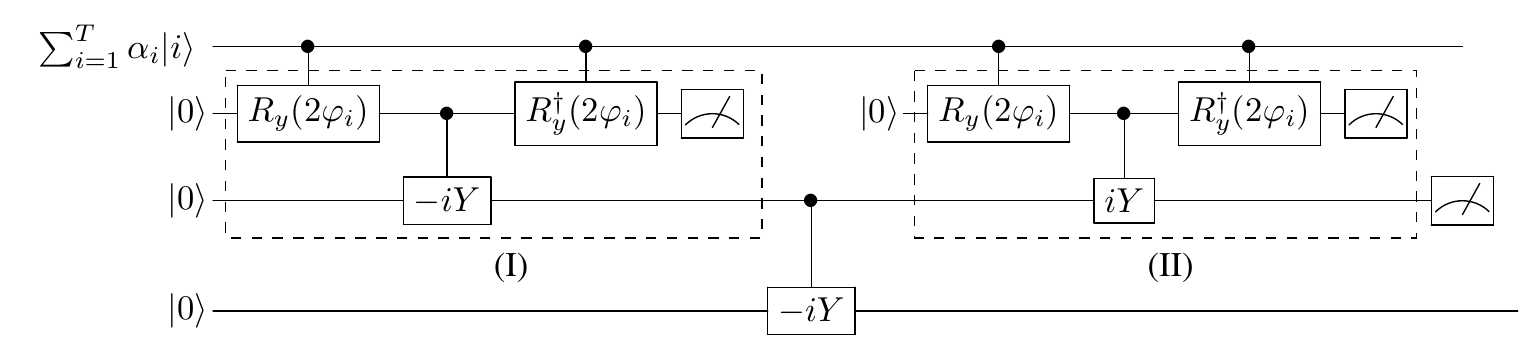}
\end{center}
\caption{Two iterations of RUS with an input controlled by a $T$-dimensional register in even superposition.}
\label{fig:rus_appendix}
\end{figure}

One could also compute the amplitudes for the state of the system in Figure \ref{fig:rus_appendix} after the bottom qubit is measured. Let $\alpha''_i=\alpha_i'\sqrt{(1-p(\varphi_i))^{s-1}p(\varphi_i)/P'_s}$. Suppose the bottom qubit is measured $|1\rangle$ and corrected for $w$ consecutive trials, and during each trial $i$, $r_i$ and $s_i$ trials were needed for circuit (I) and (II) respectively. Let $\vec{r}=(r_1,r_2,\cdots,r_w)$ and $\vec{s}=(s_1,s_2,\cdots,s_w)$. Then at the $(w+1)$-st trial the state of the system can be written as
\begin{equation}
\sum_{i=1}^T\alpha''_i\sqrt{\frac{F(\varphi_i;\vec{r},\vec{s})}{P''_w}}|i\rangle\left[\sqrt{p(q(\varphi_i))}|0\rangle R_y(q(q(\varphi_i)))|0\rangle+\sqrt{1-p(q(\varphi_i))}|1\rangle R_y(\pi/4)|0\rangle\right]
\end{equation}
where $P''_w=\sum_{i=1}^T|\alpha''_i|^2F(\varphi_i;\vec{r},\vec{s})$ is the normalization factor and the function $F$ is dependent on the history of measurement outcomes at the end of circuits for each iteration of the RUS scheme, as stored in the vectors $\vec{r}$ and $\vec{s}$:
\begin{equation}\label{eq:F}
F(\varphi;\vec{r},\vec{s}) = \prod_{i=1}^w\left[
\underbrace{(1-p(\varphi))^{r_i-1}p(\varphi)}_\text{(I)}
\underbrace{(1-p(\varphi))^{s_i-1}p(\varphi)}_\text{(II)}
(1-p(q(\varphi)))
\right].
\end{equation}
For $k$ iterations of RUS circuits (for instance the circuit in Figure \ref{fig:rus_appendix} has $k=2$), we define a new function $F_k(\varphi;\vec{y},\vec{n})$ where $\vec{y}$ and $\vec{n}$ are $k$-dimensional vectors. Each $y_i$ is the total number of successes (measurement outcome $|0\rangle$) at the $i$-th level of RUS and each $n_i$ is the total number of failures (measurement $|1\rangle$) at the $i$-th level. As an example, the function $F$ in Equation \ref{eq:F} satisfies 
\begin{equation}
F_2\left(\varphi;\vec{y}=(2,0),\vec{n}=\left(\sum_{i=1}^w(r_i+s_i-2),w\right)\right) = F(\varphi;\vec{r},\vec{s}).
\end{equation}
We then define $F_k(\varphi;\vec{y},\vec{n})$ as
\begin{equation}
F_k(\varphi;\vec{y},\vec{n})=\prod_{i=1}^m(1-p(q^{\circ i}(\varphi)))^{n_i}p(q^{\circ i}(\varphi))^{y_i}.
\end{equation}
For a particular run of $k$-iteration RUS circuit with superposition of inputs as in Figure \ref{fig:rus_appendix} for $k=2$. If the circuit successfully applies $k$ iterations, the final state of the system is
\begin{equation}
\sum_{i=1}^T\alpha_i\sqrt{\frac{F_k(\varphi_i;\vec{y},\vec{n})}{P}}|i\rangle|0\rangle^{\otimes k}R_y(q^{\circ k}(\varphi_i))|0\rangle
\end{equation}
where $P=\sum_{i=1}^T|\alpha_i|^2F_k(\varphi_i;\vec{y},\vec{n})$ is the normalization factor. The vector $\vec{y}$ is a $k$-dimensional vector such that $y_i=2^{k-i}$ and $\vec{n}$ is a vector of non-negative integers corresponding to the failure record of the particular run.

\subsection{Weights and bias setting}\label{sec:wbsetting}
As can be seen from Equation \ref{eq:k_scale}, the closeness of the initial angle $\varphi_0$ to the threshold $\pi/4$, as measured by $\Delta_0$, determines how many RUS iterations are needed for the final state to be at most $\epsilon$ away from the respective attractor. Therefore if $\Delta_0$ is allowed to be arbitrarily small, the number of RUS iterations is unbounded. To prevent this situation we restrict to neural networks where the weights and bias values can be represented in resolution $\delta$, namely 
\begin{equation}\label{eq:wb}
w=k_w\delta,\qquad
b=k_b\delta+\delta/2 
\end{equation}
for $k_w,k_b\in\mathbb{Z}$. The extra $\delta/2$ term in the bias is intended such that for any $n$, $x_i\in\{-1,1\}$ with $i\in[n]$, $|w_1x_1+\cdots+w_nx_n+b|\ge \delta/2$.

Another issue in simulating classical feedforward neural network with our quantum setup is that the activation value $\theta=w_1x_1+\cdots+w_nx_n+b$ may also be unbounded from above, while in our case we would like the input $\varphi_i$ to be restricted to $[0,\pi/2)$. Let $w_\text{max}$ and $b_\text{max}$ be the maximum possible values of $|w|$ and $|b|$ respectively. For each time the state of the current layer (which consists of $n$ qubits) is used for updating the state of the next layer, we introduce a scaling parameter 
\begin{equation}\label{eq:gamma}
\gamma=0.7\cdot\frac{1}{w_\text{max}n+b_\text{max}} 
\end{equation}
on the activation value such that the input rotation angle $\varphi$ to the neuron in the next year is contained in $[0,\pi/2)$. 

A final subtlety is the difference between the classical variables $x_i$ taking values from $\{-1,1\}$ and corresponding qubits in state $|s_i\rangle$ with $s_i\in\{0,1\}$. This can be easily addressed by the transformation $x_i=2s_i-1$. Putting these all together, we have the circuit in Figure \ref{fig:wb} for applying a rotation by angle $\varphi=\gamma(w_1x_1+\cdots+w_nx_n+b)+\pi/4$ to a qubit representing a neuron in the next layer. Hence in the feed-forward neural network of quantum neuron, the value of $\varphi>\pi/4$ iff $\theta>0$ and $\varphi<\pi/4$ iff $\theta<0$. Also $\gamma\delta/2\le|\varphi-\pi/4|\le 0.7<\pi/4$, ensuring that $\varphi$ is always restricted to $[0,\pi/2)$. 

\begin{figure}
\centering
\includegraphics[scale=1]{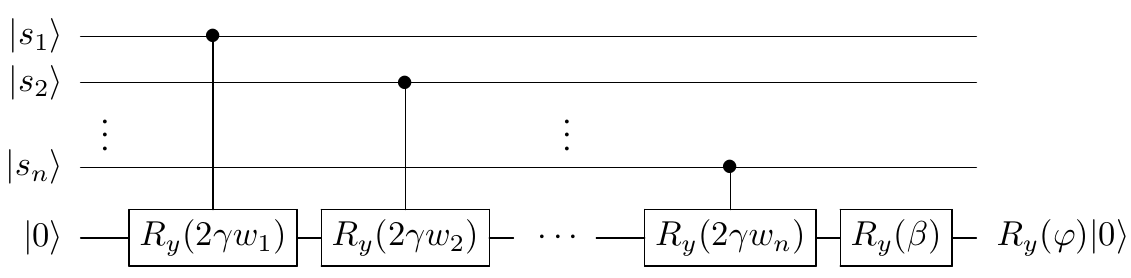}
\caption{Quantum circuit for simulating weighted and biased input process of classical neural network in Figure \ref{fig:basic_setting}a. Here the bias $\beta=\pi/4+\gamma(b-w_1-\cdots-w_n)$. The states $|s_i\rangle$ with $s_i\in\{0,1\}$ describe classical state of the neurons $x_i\in\{-1,1\}$ in the previous layer correspondingly.}
\label{fig:wb}
\end{figure}

\subsection{General property of quantum neuron}
\label{app:proof}

Here we prove Theorem \ref{thm:rustime}. Recall that $\varphi=\gamma\theta+\pi/4$ with $\theta=w_1x_1+w_2x_2+\cdots+w_nx_n+b$, $x_i\in\{0,1\}$.
Assume that the weights in $w_i$ and the bias $b$ satisfy the form described in  Appendix \ref{sec:wbsetting} (Equation \ref{eq:wb}). Then using Equation \ref{eq:k_scale} in Appendix \ref{subsec:mapq} with $\Delta=\gamma\delta/2$ and $\gamma$ as in Equation \ref{eq:gamma}, it suffices to have the number of RUS iterations as 
\begin{equation}\label{eq:k_tilde}
\tilde k=\left\lceil\log_\alpha\left(\frac{3\pi}{4\delta}\cdot\frac{1}{0.7}(nw_\text{max}+b_\text{max})\right)\right\rceil+\left\lceil\log_{1/\beta}(1/\epsilon)\right\rceil \in O\left(\log\frac{n}{\delta\epsilon^{1.52}}\right)
\end{equation}
to ensure that the following holds for any weighted input $\theta$ from ${\bf z}^{(0)}$ and $\varphi=\gamma\theta+\pi/4$:
\begin{equation}\label{eq:qg}
|q^{\circ \tilde{k}}(\varphi)-g(\theta)|\le\epsilon.
\end{equation}  
Substituting the expression for $\tilde k$ into the bound for expected runtime in Equation \ref{eq:Etk} yields a runtime of 
\begin{equation}\label{eq:prop_n1}
O(14^{\tilde k})=
O\left(
({n}/{\delta})^{\log_\alpha 14}(1/\epsilon)^{\log_{1/\beta} 14}
\right)=
O\left(
({n}/{\delta})^{2.075}(1/\epsilon)^{3.15}
\right)
\end{equation}
for the quantum neuron to produce the desired output state.

\subsection{Feedforward networks of quantum neurons}\label{app:ffnn}

Following the definitions used in Equation \ref{eq:prop}, let $|\tilde{\bf z}^{(i)}\rangle$ denote the state of quantum neurons corresponding to the $i$-th layer ${\bf z}^{(i)}$ of the classical feedforward network, $i=0,1,\cdots,\ell$. Let $|{\bf z}^{(i)}\rangle$ denote a computational basis state where each qubit is $|0\rangle$ if the corresponding bit of ${\bf z}^{(i)}$ is in state $-1$ and $|1\rangle$ if the corresponding neuron of ${\bf z}^{(i)}$ is in state $1$. 

Suppose $|\tilde{\bf z}^{(0)}\rangle=|{\bf z}^{(0)}\rangle$. Assume further that the weights in ${\bf W}^{(i)}$ and biases in ${\bf b}^{(i)}$ also satisfy the form described in Equation \ref{eq:wb}. Assuming the layer ${\bf z}^{(1)}$ (resp.\ $|\tilde{\bf z}^{(1)}\rangle$) has $m$ neurons (resp.\ qubits), then by Equation \ref{eq:prop_n1} the total runtime is $O(n^{2.075}m)$ for propagating from layer ${\bf z}^{(0)}$ to layer ${\bf z}^{(1)}$. 

Note that $\tilde k$ is a function of the number of neurons $n$ in the previous layer. Suppose each time we would like to propagate from layer ${\bf z}^{(i)}$ to layer ${\bf z}^{(i+1)}$, we use the number of RUS iterations according to Equation \ref{eq:k_tilde}. How does the error accumulate as we introduce more and more layers into the network? First of all, because of Equation \ref{eq:qg}, in $|{\bf z}^{(1)}\rangle$ each qubit has a rotation angle that is at most $\epsilon$ away from the state of its counterpart in the classical feedforward network. For example if in the classical network ${\bf z}^{(1)}=010$ then the state $|{\bf z}^{(1)}\rangle$ has at least as much overlap with $|010\rangle$ as the state
\begin{equation}\label{eq:3eps}
(\cos\epsilon|0\rangle+\sin\epsilon|1\rangle)
\otimes
(\cos(\pi/2-\epsilon)|0\rangle+\sin(\pi/2-\epsilon)|1\rangle)
\otimes
(\cos\epsilon|0\rangle+\sin\epsilon|1\rangle).
\end{equation}
The amplitude of the state $|010\rangle$ is $1-3\epsilon^2/2+O(\epsilon^4)$, while the amplitudes of the other states $|s\rangle$ is $O(\epsilon^{h(s,010)})$ with $h(x,y)$ being the Hamming distance between two bit strings $x$ and $y$. When propagating from layer $|\tilde{\bf z}^{(1)}\rangle$ to $|\tilde{\bf z}^{(2)}\rangle$, the input state to the RUS circuit is a superposition (with amplitudes concentrated on the computational basis state $|{\bf z}^{(1)}\rangle$). This leads to further complications when propagating from $|\tilde{\bf z}^{(2)}\rangle$ to $|\tilde{\bf z}^{(3)}\rangle$ and it is in general unclear whether amplitudes will remain concentrated on the states that correspond to classical neural networks. One simple way to deal with such complication is to make a measurement on the current layer after the propagation from the previous layer. Then the previous layer is always in a computational basis state and if the previous layer corresponds to the correct state that matches with the classical neural network, amplitude will always be concentrated on the correct state in the current layer.

Suppose one is propagating from the $(j-1)$-st layer to the $j$-th layer and $|\tilde{\bf z}^{(j-1)}\rangle=|{\bf z}^{(j-1)}\rangle$. Then if we measure $|\tilde{\bf z}^{(j)}\rangle$, since $\langle {\bf z}^{(j)}|\tilde{\bf z}^{(j)}\rangle=\cos^n\epsilon\ge 1-n\epsilon^2/2$, the probability of getting $|{\bf z}^{(j)}\rangle$ is $|\langle {\bf z}^{(j)}|\tilde{\bf z}^{(j)}\rangle|^2\ge 1-n\epsilon^2$. The probability of failure is then at most $n\epsilon^2$. If we let $\epsilon\le\frac{1}{2\sqrt{n}}$ then the failure probability is at most 1/4. If we make $M$ repetitions of the propagation from $|\tilde{\bf z}^{(j-1)}\rangle$ to $|\tilde{\bf z}^{(j)}\rangle$ and record the measurement outcomes, the the probability that the majority of the repetitions return a wrong state is bounded from above by Chernoff inequality. The Chernoff inequality states that for independent and identically distributed 0-1 variables $X_1$ through $X_M$, where each $X_i$ has $1/2+\xi$ probability of being 1 and $1/2-\xi$ probability of being 0, the probability that $X_1+\cdots+X_M\le m/2$ is at most $e^{-2\xi^2M}$. If we use the identification $\epsilon=\frac{1}{2\sqrt{n}}$ and further restrict the failure probability to be within some tolerance $\eta$, we have the minimum number of repetitions being $M\ge 8\ln(1/\eta)$. For $\eta=10^{-9}$, namely one in a billion, the number of repetitions ``merely'' has to be at least 166. After $M$ repetitions are executed and the majority string is identified, we run the propagation again to obtain the majority string. This extra runs should contain only $1/|\langle {\bf z}^{(j)}|\tilde{\bf z}^{(j)}\rangle|^2\le 1+n\epsilon^2 + \cdots\in O(1)$ repetitions on average.

Once we prepare the state $|{\bf z}^{(j)}\rangle$, we could continue the same procedure for propagating from the $j$-th layer to the $(j+1)$-st layer. The total number of applications of $k$-iteration RUS circuit needed for propagating from $|\tilde{\bf z}^{(0)}\rangle$ to $|\tilde{\bf z}^{(\ell)}\rangle$ is then $O(nM\ell)$ with $n$ being the \emph{maximum} layer size. The probability that the majority outcome of measuring $|\tilde{\bf z}^{(i)}\rangle$ is the same as the corresponding layer $|{\bf z}^{(i)}\rangle$ in the classical network is then $(1-\eta)^\ell$, with $\eta$ being the error probability for the propagation of a single layer to the next. The total error probability is then $\nu=1-(1-\eta)^\ell=O(\eta\ell)$. 
Then combining Equation \ref{eq:prop_n1} and the above estimate yields the time cost for simulating, with success probability at least $1-\nu$ and error at most $\epsilon$, an $\ell$-layer classical deep feedforward neural network with layer size at most $n$ and weights/bias setting described in Equation \ref{eq:wb}: 
\begin{equation}\label{eq:time_complexity}
O\left(\frac{n^{3.075}\ell}{\delta^{2.075}\epsilon^{3.15}}\log\left(\frac{\ell}{\nu}\right)\right).
\end{equation}
The total number of qubits needed for the simulation can then be characterized as
\begin{equation}
O\left(n\ell+\log\frac{n}{\delta\epsilon^{1.52}}\right),
\end{equation}
which is almost linear in the number of neurons in the classical feedforward neural network. From Equation \ref{eq:time_complexity} we see that if the number of layers is constant, then assuming the other error parameters $\nu$, $\delta$ and $\epsilon$ are constant, the quantum neural network runs in roughly $O(n^3\ell)$ time, which has only a linear overhead compared with the classical case, which requires $O(n^2\ell)$ computation due to the bipartite structure of how each layer is connected to its next layer.

\subsection{Hopfield networks of quantum neurons}\label{app:hopfield}

Following the discussion in Appendix \ref{app:ffnn}, for a Hopfield network whose weights and bias parameters satisfy Equation \ref{eq:wb}, we could consider the same RUS circuit construction for simulating each update. Appendix \ref{app:ffnn} gives the runtime estimate for a single update. We repeat the circuit several times and measure the output qubit each time. We then identify the majority state (0 or 1) as the new state for the neuron, and change the state of the corresponding qubit in the network accordingly. Therefore we do not need the linear qubit overhead, since we could always recycle the output qubit for the next update. By the Chernoff argument presented in Appendix \ref{app:ffnn}, to ensure error probability within $\eta$ we need $O(\log(1/\eta))$ repetitions. For $t$ updates the error probability is $O(\eta t)$. Putting these together, to simulate $t$ updates of an $n$-neuron Hopfield network with weights and biases satisfying Equation \ref{eq:wb}, up to error $\epsilon$ and success probability at least $1-\nu$, the expected runtime is
\begin{equation}
O\left(\frac{n^{2.075}t}{\delta^{2.075}\epsilon^{3.15}}\log\left(\frac{t}{\nu}\right)\right).
\end{equation}
The total number of qubits needed for the simulation is
\begin{equation}
O\left(n+\log\frac{n}{\delta\epsilon^{1.52}}\right).
\end{equation}

\end{widetext}

\end{document}